\documentclass[a4paper,12pt,oneside,final,openbib]{article}
\pdfoutput=1
\usepackage[utf8]{inputenc}
\usepackage{url}
\usepackage{cite}
\usepackage{hyperref}
\usepackage{breakurl}
\usepackage{graphicx}
\usepackage{subcaption}
\usepackage{multirow}

\usepackage{amsmath,amssymb,epsfig}
\usepackage{color}

\usepackage{xcolor}

\begin{document}

%%%%%%%%%%%%%%%%%%%%%%%%%%%%%%%%%%%%%%%%%%%%%%%%%%%%%
\title{Analytically Consistent Reconstruction of Finite Data Using Padé Sequences}

%Padé Approximants as Noise Filtering for Experimental Data.// Analytic consistency, Padé approximants as an analytic reconstruction method in presence of noisy experimental data.//Recorvering Analytic Consistency by reconstructing Finite Data using Padé Sequences//Analytic Consistency from Experimental Data using Padé Sequences

\author{
% No sé si quieres poner Emerson Miguel o solo Emerson
Emerson Díaz, Balma Duch, 
Pere Masjuan 
  \\
  {\normalsize \em
  Grup de F\'{\i}sica Te\`orica, Departament de F\'{\i}sica, 
  Universitat Aut\`onoma de Barcelona,}
  \\
  {\normalsize \em
  and Institut de F\'{\i}sica d'Altes Energies (IFAE), }
  \\
  {\normalsize \em
  The Barcelona Institute of Science and Technology (BIST), }
  \\
  {\normalsize \em
  Campus UAB, E-08193 Bellaterra (Barcelona), Spain}
}

\date{\today}

\maketitle

\begin{abstract} 

Reconstructing the analytic structure of a function from finite datasets is a fundamental problem across theoretical, numerical, and experimental physics. While Padé approximants provide a natural framework, finite-information effects, as well as statistical and systematic uncertainties, may obscure the underlying analytic structure and limit reconstruction reliability. In this work, we reinterpret the appearance of Froissart doublets not merely as numerical artifacts but as \textit{diagnostic objects} carrying information about the analytic consistency of the input data. Accordingly, we develop a general Padé-based algorithm that exploits the dynamics of Froissart doubltes along Padé sequences to identify localized inconsistencies and iteratively reconstruct the analytic structure most compatible with the data. The method requires no model for the origin of the inconsistencies and distinguishes genuine analytic features from spurious structures induced by finite-information effects. We validate it using Stieltjes functions, realistic pseudo-experimental datasets with statistical and systematic uncertainties, and general holomorphic functions. The complete algorithm is provided as a supplementary Mathematica notebook in an open GitLab repository.

\end{abstract}

Keywords: Analyticity, reconstruction, experimental data, lattice data, Padé sequences

\clearpage

%\pacs{13.35.Dx, 14.65.-q, 11.55.Hx, 12.38.Bx.}

%%%%%%%%%%%%%%%%%%%%%%%%%%%%%%%%%%%%%%%%%%%%%%%%%%%%%
\section{Introduction}
\label{sec:Introduction}

Many problems in physics require reconstructing the analytic structure of an underlying function from finite and imperfect datasets. Such situations arise naturally in experimental measurements, lattice field-theory simulations, and truncated perturbative expansions, where only a finite number of observations or expansion coefficients is available and each is affected by finite precision and/or statistical uncertainties. Moreover, the experimental or numerical procedure itself may introduce localized inconsistencies or systematic distortions that mimic genuine physical structures. Although the underlying observables often satisfy well-established analytic constraints, these effects can obscure their analytic behavior, making it difficult to distinguish genuine physical features and analytic structures from artifacts generated by the data. The central challenge is therefore to recover the underlying function together with
its analytic structure encoded in finite datasets while separating the underlying
physical content from spurious contributions~\cite{Abbott:2026wdw}.

Padé approximants (PAs) provide a natural framework for addressing this reconstruction problem because they incorporate analytic information directly into the approximation. Unlike polynomial expansions, they naturally reproduce nontrivial analytic structures, including poles and branch cuts, from finite information, making them particularly effective for analytic continuation and extrapolation \cite{baker1996pade}. When constructed from finite experimental or numerical datasets, however, their behavior is governed not only by the underlying function but also by the consistency of the available information. Beyond a certain approximation order, the additional degrees of freedom cease to reveal new information about the underlying function and instead become increasingly sensitive to finite-information effects, localized inconsistencies, and systematic distortions present in the dataset. This transition is manifested by the appearance of spurious pole-zero pairs, known as Froissart doublets \cite{froissart1969pade}, whose properties provide a direct probe of the analytic consistency of the data.

The appearance of Froissart doublets is widely regarded as the hallmark of the breakdown of Padé reconstruction due to finite-information effects. Rather than being viewed solely as numerical artifacts, however, these structures may also be regarded as \textit{diagnostic objects} that encode valuable information about the consistency of the underlying dataset. This interpretation is consistent with a substantial body of work showing that Froissart doublets naturally emerge in PAs constructed from noisy, incomplete, or finite-precision data. Their statistical properties, asymptotic behavior, and implications for analytic continuation have been investigated. They were first identified as spurious pole-zero pairs appearing in rational approximations by Froissart \cite{froissart1969pade}. Their role was later revisited by Bessis, who showed that these structures can carry information about noise affecting the input data \cite{Bessis:1996}. This observation motivated two complementary directions: improvement of the numerical stability of PAs by identifying and removing spurious structures through techniques such as singular-value decomposition or explicit pole-zero cancellation \cite{Gonnet2013,Beckermann2018}; investigation of the information carried by Froissart doublets, showing that their statistical properties and distribution in the complex plane can reveal characteristics of noisy data and weak signals \cite{Gilewicz1997,Gilewicz2003,Perotti2010,YamadaKensuke,Boito:2022rad,Falcao:2022gxt}. Despite these advances, Froissart doublets have been used primarily as indicators of the limitations of Padé approximation\cite{Binosi:2019ecz,MasjuanQueralt:2010hav}. Their potential as quantitative probes of data consistency —and, more importantly, as guides for reconstructing the underlying analytic structure— has remained largely unexplored. This perspective forms the basis of the present work. Rather than treating Froissart doublets as undesirable by-products of rational approximation, we exploit their dynamics along Padé sequences as a source of information about the dataset itself. As we show, these dynamics identify localized inconsistencies associated with statistical fluctuations or systematic distortions, providing the key ingredient for reconstructing the analytic structure underlying finite data.

Building on these ideas, we develop a general framework that transforms the diagnostic information contained in Froissart doublets into a practical reconstruction algorithm. The method iteratively exploits the dynamics of Padé sequences to identify localized inconsistencies in finite datasets and recover the analytic structure that is compatible with the available information. Because it relies only on the analytic properties of the approximants, the procedure is largely independent of the origin of the data and does not require prior assumptions about the form of the underlying noise or systematic effects. We demonstrate its performance on Stieltjes functions, realistic pseudo-experimental datasets affected by statistical and systematic uncertainties, and more general holomorphic functions. To facilitate its application, the complete implementation is provided as supplementary  Mathematica notebook in an open GitLab repository.

The paper is organized as follows. Section~\ref{sec: General Formalism} reviews Padé approximants for Stieltjes functions and discusses their ability to analytically  reconstruct the target function from finite datasets through fitting procedures. Section~\ref{sec:Method} introduces the analytical reconstruction method's algorithm based on the dynamics of Froissart doublets along Padé sequences. Sections~\ref{sec:Validation in Controlled Settings with Discrete Noisy Data}, ~\ref{sec:Analysis in a Realistic Scenario} and  ~\ref{sec:Extension to holomorphic functions}  present several representative applications of the proposed method. We end with conclusions in Sec.~\ref{sec:conclusion} and relegate implementation details to two appendices, which describe in detail the user-friendly Mathematica notebook and the openly available GitLab repository. 

%%%%%%%%%%%%%%%%%%%%%%%%%%%%%%%%%%%%%%%%%%%%%%%

\section{Padé approximants to Stieljes Functions: Analytically Consistent Reconstruction Method}
\label{sec: General Formalism}

PAs \cite{baker1996pade} provide a natural framework to address this problem. They are approximant rational functions to a given function $f(z)$ constructed from a finite number of coefficients $f_n$ of its power series $f(z)=\sum_{n=0}^{\infty} f_n z^n$ via a matching-through-order set of conditions. More explicitly, a PA $P_N^M(z)=R_N(z)/Q_M(z)$, with $R_N(z)$ and $Q_M(z)$ polynomials of degree $N,M$ respectively, and with $Q_M(0)=1$ without lose of generality, satisfies
\begin{equation}
f(z)-P_N^M(z)=\mathcal{O}(z^{N+M+1}).
\end{equation}
As such, PAs can be interpreted as nonlinear analytic continuations of truncated power series, often reproducing global analytic features of a function such as poles and branch cuts and remaining accurate beyond the radius of convergence of the original expansion. This makes them a powerful tool in quantum field theory, statistical physics, and numerical analysis.

A particularly important class of functions is given by Stieltjes functions. In this case, strong convergence theorems ensure that PAs constructed from the Taylor expansion converge to the exact function in the limit of infinite order \cite{baker1996pade}. Moreover, diagonal and sub-diagonal approximants satisfy ordering and bounding properties with respect to their neighbors, which induces a natural hierarchy of approximations that constrains the systematic error within the Padé sequence \cite{Peris:2006ds,Masjuan:2009wy,Boito:2018rwt,Boito:2021scm,Boito:2024yat}.

The above results rely crucially on the availability of exact Taylor coefficients and the associated analytic structure \cite{Masjuan:2007ay,Masjuan:2008fr,Masjuan:2008cp}. In practical applications, however, one is often confronted with situations where such coefficients are not directly accessible, and PAs must instead be constructed from finite and potentially noisy data through fitting procedures \cite{Masjuan:2008cp,Masjuan:2008fv,Masjuan:2009wy,Masjuan:2012wy,Escribano:2013kba,Masjuan:2015lca,Escribano:2015nra,Escribano:2015yup,Masjuan:2015cjl,Masjuan:2017tvw,Gonzalez-Solis:2018ooo,Gonzalez-Solis:2021pyh,Boito:2024yat,Vaziri:2025rfq}.

In this latter setting, the convergence properties established in the analytic framework, including those associated with the textbook of Stieltjes functions, are no longer guaranteed. In addition, the presence of noise in such data sets further blurs the notion of convergence and its convergence velocity. Therefore, in these more realistic cases one must adapt the convergence convention to the finite-data problem.

In this work, convergence is thus understood as both the ability of the Padé sequence to properly fit the data \textit{and} to separate the underlying function from the effective noise introduced by the finite information content of the dataset with more and more accuracy. 

Even when the data are generated exactly from a known function, finite sampling density and finite numerical precision introduce an effective source of uncertainty. As the Padé order is increased, the sequence eventually reaches a regime where additional degrees of freedom no longer encode further information about the underlying function, but instead describe these finite-information effects. This results in the appearance of spurious pole-zero pairs known as Froissart doublets \cite{baker1996pade}. These structures are transient in nature and thus unstable under variations of the Padé order and do not correspond to genuine analytic features of the function.

Similar phenomena have been observed when PAs are constructed from noisy Taylor coefficients \cite{baker1996pade,Costin:2022}. The appearance of Froissart doublets in this case motivated the authors of Ref.\cite{Costin:2022} to  define a critical Padé order, denoted by $N_c$. For orders below $N_c$, increasing the Padé order improves the convergence to the original function, as the additional parameters allow the sequence to capture further information about it. The sense of convergence in this case, with PAs built from Taylor expansion coefficients, follows the canonical definition \cite{baker1996pade}. Beyond this $N_c$ threshold, additional poles and zeros no longer provide useful information about the underlying analytic structure, and thus about the original function, and instead reflect the finite resolution of the dataset. Accuracy and precision of the Taylor coefficients must be retained in order to understand the appearance of Froissart doublets as both may generate them.  Accuracy is obviously important: if a given coefficient $f_n$ contains a distortion $\epsilon$, the PA would as well be distorted by developing a Froissart doublet. Precision, nonetheless, plays a similar role, since the construction of a PA of order $N$ from the matching-through-order conditions demands the knowledge of the coefficients with increasing precision, roughly with $2N$ digits. Lack of precision also results in the appearance of Froissart doublets. These two doublets have, however, different nature and different pattern within a given PA sequence \cite{Binosi:2019ecz,Costin:2022}.

Dealing with datasets, the situation and the corresponding Froissart analysis is more complicated. In this case, the critical order $N_c$ is controlled by both the numerical precision and the density $\rho$ of the data, the number of data points per unit interval. If noise is present in the dataset, increasing the precision allows the Padé sequence to resolve higher-order information before entering the noise-dominated regime. Precision $\epsilon$ here is understood as at which particular digit the datum is truncated before leaving an infinite set of zeros after it. Likewise, increasing the sampling density delays the appearance of Froissart doublets by providing a more detailed representation of the function. While the distinction of the two effects can be observed in analytic exercises as we aforementioned, cf. \cite{Costin:2022}, for numerical fit procedures the entanglement of both are so involved that separating them is way too difficult. This analytic separation is not going to be pursued in this work.  

Nevertheless, we can illustrate this intricated mechanism with an example. We consider data generated from the function $f(z)=\log(1+z)/z$, uniformly sampled over the interval $z\in[0,10]$ with density $\rho=2.5$ (25 points) and truncated to a fixed numerical precision $\epsilon$. Diagonal PAs, $P_N^N$, are then constructed via a fitting procedure from the resulting dataset. As shown in Table~\ref{tab:precision_Nc}, the critical order $N_c$ increases systematically with the numerical precision $\epsilon$ of the input data, as the sequence can resolve higher-order information (higher derivatives) before truncation errors induce the formation of Froissart doublets. An analogous trend is found when increasing the sampling density $\rho$, although in that case the limiting factor is the finite spatial resolution of the dataset rather than numerical precision. The nonlinearity of the $N_c$ versus $\rho$ and $\epsilon$ prevents, unfortunately, a generic study and thus goes beyond the scope of the present work, as aforementioned. 

%%%%%%%%%%%%%%%%%%%%%%%%%%%%%%%%%%%%%%%%%%%%%%%%%

\begin{table}[h]
\caption{Critical PA order $N_c$ obtained from diagonal approximants $P_N^N$ constructed from samples of $f(z)=\log(1+z)/z$ with data truncation precision $\epsilon$ and sampling density $\rho$.}
\label{tab:precision_Nc}
\centering
\begin{tabular}{ccc}
\hline
Precision ($\epsilon$) with $\rho =2.5$ & Critical Order ($N_c$) & Density ($\rho$) with $\epsilon = 10^{-8}$ \\ \hline
$10^{-3}$ & 2 & 0.5 \\
$10^{-4}$ & 3 & 0.7 \\
$10^{-6}$ & 4 & 1   \\
$10^{-8}$ & 5 & 2.5 \\ \hline
\end{tabular}
\end{table}
%%%%%%%%%%%%%%%%%%%%%%%%%%%%%%%%%%%%%%%%%%%%%%%%%

The structure of Froissart doublets in noisy Padé reconstructions has been studied extensively in the literature, either based on Taylor expansions or datasets, as discussed in the Introduction. Of particular interest is Ref.\cite{Bessis:1996} who used the Froissart-doublets structure generated by noisy-Taylor coefficients and the results of Refs  \cite{Gammel1,Gammel2,GilewiczTruongvan} to analyze a particular PA (the $P^6_7$ in that case, not a sequence) fitted to a dataset generated from a single-pole function. Ref.\cite{Bessis:1996} observed then that noise does not distribute uniformly in the complex plane, but instead organizes into localized pole-zero structures. Based on this analysis, which is subject to specific assumptions, poles and zeros can be separated into three distinct families: (i) pole-zero pairs associated with numerical roundoff errors, (ii) pole-zero pairs induced by noise in the input data, and (iii) poles and zeros compatible with the genuine analytic structure of the underlying function. Ref.\cite{Bessis:1996} discarded the first two families, obtaining a filtered reconstruction, an expected $P^1_1$ that better reflects the genuine analytic structure of the target function, \textit{better} compared to the original fitted $P^6_7$ as the outcome has the known analytic function used to generate the data. 

In this work, we build on this observation but we are forced to adopt a different viewpoint since in real cases a) the analytic structure of the function may not be known b) the noise should not be homogeneously distributed c) dataset will contain both statistic and systematic uncertainties d) noisy data still contain reliable information which should be preserved. In the finite-data setting considered here, the separation between the different sources of noise is not operationally well-defined. Finite sampling density $\rho$ and finite numerical precision $\epsilon$ produce the same observable effect along the Padé sequence: unstable Froissart doublets, whose positions depend sensitively on the Padé order. Consequently, these effects cannot be distinguished solely from the resulting approximants and must be regarded as a common finite-information contribution.

We therefore classify the poles and zeros according to their behaviour along the Padé sequence, leading to a classification into three families: (i) \underline{unstable} Froissart doublets associated with finite-information effects, (ii) \underline{recurrent} Froissart doublets associated with inconsistencies in the input data, and (iii) poles and zeros \underline{compatible} with the analytic structure of the underlying function.

The distinction between the first two classes is not causal but dynamical and, thus, observed from the PA sequence. Doublets in class (i) exhibit strongly unstable locations as the Padé order is varied, whereas those in class (ii) recur in similar regions of the complex plane across successive approximants. Such recurrence provides evidence for localized inconsistencies, as these structures persist under changes of approximation order and cannot be interpreted as transient finite-information effects. This observation provides the basis for the analytical reconstruction strategy developed in the next section.

\section{Analytic Reconstruction Method}
\label{sec:Method}

Our method filters noise in datasets by employing the analytic properties of Stieltjes functions through a sequence of diagonal PA. Its advantage is that the method returns actually an algorithm which is doable, scalable, transportable, and access free from our \textcolor{red}{\href{https://gitlab.pic.es/bduch/reconstruction-finite-data-pade-sequences}{GitLab}} repository.
The goal of the algorithm is to split, within each PA, which parts converge to the Stieltjes function and which parts to the noised fraction. Given a set of discrete data points (assumed to lie on the real axis), the algorithm iteratively identifies and corrects data points that deviate from their expected analytic behavior. The input of the method is therefore a discrete dataset affected by noise (understood as both external noise and precision round-off), while the output is a corrected dataset defined on the same sampling grid and with similar numerical precision.

The procedure begins by constructing a sequence of diagonal PAs, $P_N^N(x)$, up to a given N, and fitting them to the dataset. We define a grid node as each coordinate point at which a data value is available. In order to analytically reconstruct data points, we need to identify noisy subdatasets. Then, for each approximant, the poles of the rational function are calculated, and only those that lie within the data domain and within a predefined tolerance (see Appendix~\ref{app:Conventions} for details) of a grid node are retained. Each retained pole is then assigned to its nearest sampling node, and a voting system tracks how frequently each node hosts a recurring pole across the sequence of approximants.

Through the PA sequence, persistent retained poles would reveal localized deviations from the analytic structure expected of a Stieltjes function, and these points are consequently prioritized for correction. To define this process, each PA can be conceptually decomposed into two components:
\begin{equation}
\label{ec: PAStieltjesNoise}
P_N^N(x) = P_M^M{}_{\mathrm{Stieltjes}}(x) + P_{N-M}^{\,N-M-1}{}_{\mathrm{Noise}}(x),
\end{equation}
\noindent
where \(P_M^{M}{}_{\mathrm{Stieltjes}}(x)\) satisfies the properties of a Stieltjes function \footnote{In particular, its pole-zero distribution tends to accumulate along the branch cut of the function in the complex plane.}, and \(P_{N-M}^{\,N-M-1}{}_{\mathrm{Noise}}(x)\) represents noise arising from experimental measurements, numerical round-off, and data precision limitations. The Stieltjes component of highest order M is chosen as the reference for dataset correction, and if multiple candidates share this maximal order, the one minimizing the mean absolute error with respect to the data is selected.

Via spliting $P_M^M{}_{\mathrm{Stieltjes}}(x)$ from $P_{N-M}^{\,N-M-1}{}_{\mathrm{Noise}}(x)$, corrections are applied iteratively to the highest-ranked nodes -those most frequently accumulating poles-, adjusting their values toward the predictions of the selected Stieltjes approximant at that point, with each modification accepted only if it preserves the positivity and convexity properties characteristic of Stieltjes functions (cf. Ref.\cite{baker1996pade}). As the dimension of the subdata set presenting noise is not known in advance, we focus the corrections point by point iteratively. To monitor the improvement of the dataset, we use the Mean Absolute Error (MAE) defined as the average absolute difference between the dataset values and the exact analytic function evaluated at the same nodes.

After each single update, the full PA sequence is recomputed and the pole analysis is repeated, with iterations continuing until either no persistent poles remain or successive corrections become negligible; or the leading candidates fall below a prescribed vote threshold (which may be set to zero). By the end of the procedure, the dataset is reconstructed from systematic errors while maintaining the analytic structure characteristic of a Stieltjes function.

Next subsection explore these features with a set of controlled examples.

%%%%%%%%%%%%%%%%%%%%%%%%%%%%%%%%%%%%%%%%%%%%%%%%%%%%%
\subsection{Validation in Controlled Settings with Discrete Noisy Data}
\label{sec:Validation in Controlled Settings with Discrete Noisy Data}

To assess the performance of the proposed reconstructing algorithm, we consider controlled numerical examples in which the analytic form of the underlying function is known exactly. This allows a direct comparison between noisy observations, analytic reconstructions, and the corresponding reference solution, and thus a validation of the method.

All datasets considered in this section are generated with numerical precision $10^{-4}$. A higher-precision case ($10^{-6}$) is included to assess the stability of the method with respect to numerical accuracy.

All examples are based on samples of the Stieltjes-type function
\begin{equation}
\label{ec:stieltjesfunctionLog}
f(z) = \frac{\log(1+z)}{z},
\end{equation}
which has a branch point at $z=-1$ and a branch cut along the negative real axis. The function is sampled over the interval $z \in (0,10]$ using a uniform discretization, and different density $\rho$. From each dataset, a subset of $n$ sampling nodes is randomly selected and perturbed by multiplicative noise of relative amplitude up to $20\%$, according to
\begin{equation}
\tilde f(z_i) = f(z_i)\,(1+\epsilon_i), \qquad \epsilon_i \sim \mathcal{U}(-0.2,0.2).
\end{equation}
The resulting values $\tilde f(z_i)$ constitute the input to the analytic reconstruction algorithm, while the output corresponds to the reconstructed dataset obtained after applying the noise reconstruction procedure.

%%%%%%%%%%%%%%%%%%%%%%%%
\begin{figure}[!htbp]
\centering
%\hspace{0.5 cm}
\begin{subfigure}[b]{0.49\textwidth}
        \includegraphics[width=\textwidth]{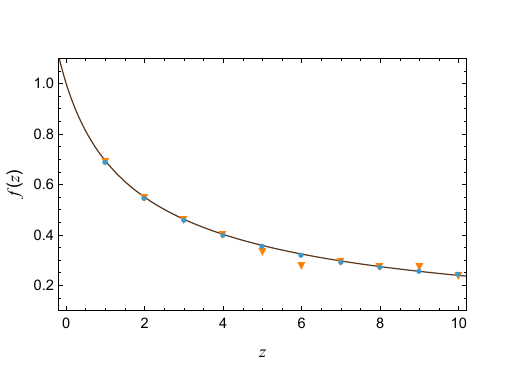}
        \caption{Subset with $\rho = 1$ and 3 noisy data points}
        \label{3outof10}
\end{subfigure}
\begin{subfigure}[b]{0.49\textwidth}
        \includegraphics[width=\textwidth]{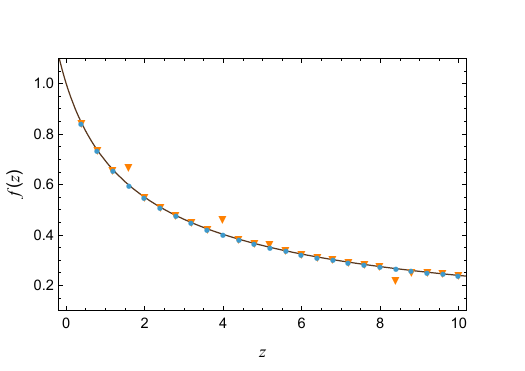}
        \caption{Subset with $\rho = 2.5$ and 5 noisy data points}
        \label{5outof25}
\end{subfigure}
%\hspace{0.5 cm}
\begin{subfigure}[b]{0.49\textwidth}
        \includegraphics[width=\textwidth]{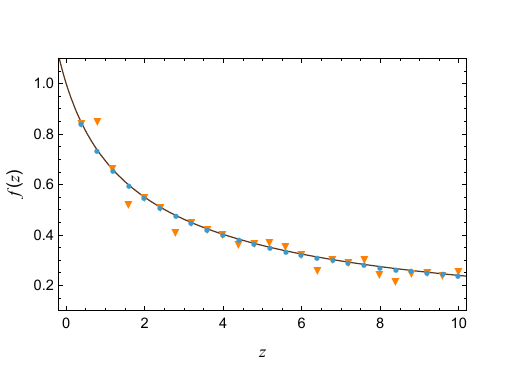}
        \caption{Subset with $\rho = 2.5$ and 15 noisy data points}
        \label{15outof25}
\end{subfigure}
%\hspace{0.5 cm}
\begin{subfigure}[b]{0.49\textwidth}
        \includegraphics[width=\textwidth]{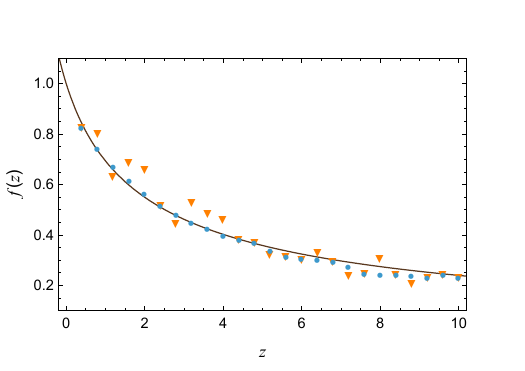}
        \caption{Subset with $\rho = 2.5$ and 25 noisy data points}
        \label{25outof25}
\end{subfigure}
\begin{subfigure}[b]{0.49\textwidth}
        \includegraphics[width=\textwidth]{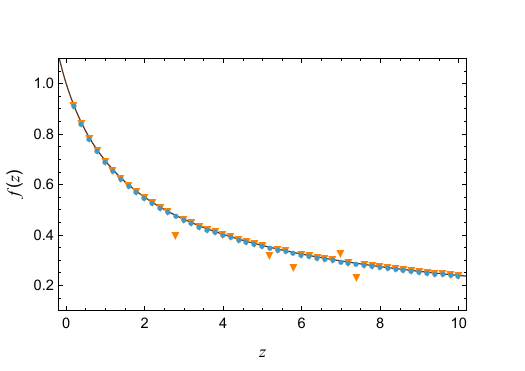}
        \caption{Subset with $\rho = 5$ and 5 noisy data points}
        \vspace*{3mm}
        \label{5outof50}
\end{subfigure}
\begin{subfigure}[b]{0.49\textwidth}
        \includegraphics[width=\textwidth]{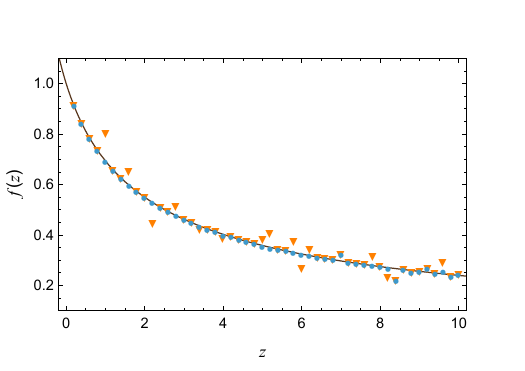}
        \caption{Subset with $\rho = 5$ and 20 noisy data points}
        \vspace*{3mm}
        \label{20outof50}
\end{subfigure}
\caption{
Representative data subsets of the Stieltjes function in Eq.~\ref{ec:stieltjesfunctionLog}, generated under different noise ($n=3,5,15,20,25$) and sampling configurations ($\rho = 1, 2.5, 5$) with numerical precision $10^{-4}$. 
The black continuous curve corresponds to the exact analytic function. 
Orange triangles denote the input data after randomly perturbing $n$ sampling points by additive noise of amplitude up to $20\%$. 
Blue dots represent the analytic reconstruction output produced by the Padé-based algorithm. 
Improvement ranges from $\sim 60\%$ to  $\sim 99\%$.}
\label{fig:Exampleswithprecision10_4}
\end{figure}
%%%%%%%%%%%%%%%%%%%%%%%%%%%%%%%%%%%

Figure~\ref{fig:Exampleswithprecision10_4} shows representative results for different sampling densities ($\rho=1,2.5,5$) and noise configurations $n$. The reconstructed outputs generally lie closer to the reference analytic function, with the degree of improvement depending on both the sampling density and the distribution of the perturbed points. For low sampling density ($\rho=1$, Fig. \ref{3outof10}), the reconstruction is more sensitive to noise, although a clear reduction of the perturbations is still observed, with a relative improvement of about 78$\%$. For intermediate and higher densities ($\rho=2.5$, Figs. \ref{5outof25}, \ref{15outof25}, \ref{25outof25}  and $\rho=5$, Figs. \ref{5outof50}, \ref{20outof50}), the method typically achieves a more accurate recovery of the underlying structure, with the final outcome depending on the number and location of the noised data points, ranging from 62$\%$ when all data points are noisy, up to $\sim 99\%$ with 5 noisy data point over 25 or over 50.

%%%%%%%%%%%%%%%%%%%%%%%%%%%%%%%%%%%%%
\begin{table}[h!]
\caption{Mean Absolute Value (MAE) before and after applying the analytic reconstruction algorithm and the corresponding relative improvement for different noise configurations and sampling densities ($\rho$). Results correspond to the data subsets shown in Fig.~\ref{fig:Exampleswithprecision10_4}.}
\centering
\label{tab:MAE_filtering}
\begin{tabular}{ccccc}
\hline
                            & Total Number of   & MAE                  & MAE                  & Relative           \\
                            & Noisy Data Points & Before Reconstruction     & After Reconstruction      & Improvement $(\%)$ \\ \hline
$\rho=1$                    & 3                 & $8.81 \cdot 10^{-3}$ & $1.95 \cdot 10^{-3}$ & $77.87$            \\ \hline
\multirow{3}{*}{$\rho=2.5$} & 5                 & $7.88 \cdot 10^{-3}$ & $8.0 \cdot 10^{-6}$  & $99.90$            \\
                            & 15                & $2.04 \cdot 10^{-2}$ & $4.40 \cdot 10^{-4}$ & $97.85$            \\
                            & 25                & $3.48 \cdot 10^{-2}$ & $1.31 \cdot 10^{-2}$ & $62.31$            \\ \hline
$\rho=5$                    & 5                 & $5.09 \cdot 10^{-3}$ & $1.20 \cdot 10^{-5}$ & $99.76$            \\
                            & 20                & $1.50 \cdot 10^{-2}$ & $2.87 \cdot 10^{-3}$ & $80.83$            \\ \hline
\end{tabular}
\end{table}
%%%%%%%%%%%%%%%%%%%%%%%%%%%%%%%%%%%%%5

These observations are quantified in Table~\ref{tab:MAE_filtering}, where the MAE is reported before and after reconstruction where a reduction of the error is observed in all cases.
%, although the magnitude of the improvement varies across configurations: Cases with few perturbed points (e.g. $n=5$) show the largest relative gains, while configurations with a higher number of noisy samples (e.g. $n=20$ at $\rho=5$ or $n=25$ at $\rho=2.5$) exhibit more moderate but still significant improvements.}

Beyond point-wise accuracy, the analytic reconstruction procedure also modifies the pole structure obtained from the sequence of PAs. After reconstructing, the Stieltjes component $P_M^M{}_{\mathrm{Stieltjes}}(x)$ becomes more persistent across the sequence, remaining stable over a wider range of approximation orders, whereas the noise component $P_{N-M}^{\,N-M}{}_{\mathrm{Noise}}(x)$ progressively loses persistence and is confined to lower-order contributions. Sequences here explored with $\epsilon = 10^{-4}$ ($N_c=3$ in all cases) run up to $M+N=12$, with initial value at most $M\leq2$ (most commonly $M=1$), and reaching up to $M=3$ after the analytic reconstruction algorithm in the cases exhibiting the largest relative improvement (typically above $\sim 90\%$). This pole-position shift reflects the effect of the iterative correction scheme, which selectively suppresses unstable pole contributions and reinforces those compatible with the analytic structure of the data.

%%%%%%%%%%%%%%%%%%%%%%%%
\begin{figure}[h!]
\centering
\includegraphics[width=0.6\textwidth]{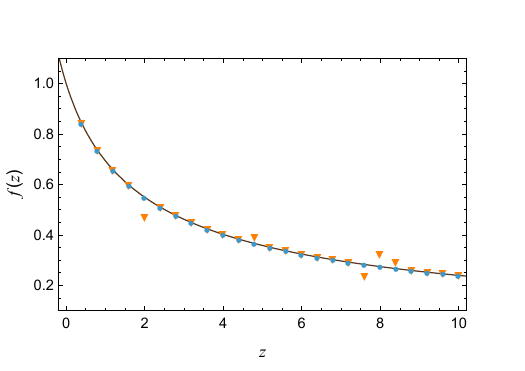}
\caption{
Representative data subset of the Stieltjes function in Eq.~\ref{ec:stieltjesfunctionLog}, generated under sampling density $\rho = 2.5$ with $n=5$ noisy data points and numerical precision $10^{-6}$. The black continuous curve corresponds to the exact analytic function. Orange triangles denote the input data after randomly perturbing $n$ sampling points by multiplicative noise of amplitude up to $20\%$. Blue dots represent the analytic reconstruction output produced by the Padé-based  algorithm.}
\label{fig:precision_1e6}
\end{figure}
%%%%%%%%%%%%%%%%%%%%%%%%%%%%%%%%%%%

We additionally test the algorithm under higher numerical precision ($10^{-6}$), considering the case $\rho=2.5$ with $n=5$ perturbed samples, as illustrated in Fig.~\ref{fig:precision_1e6}. In this regime, the method correctly identifies and removes all noise contributions, achieving a relative improvement of $97.91\%$, comparable to that observed at precision $10^{-4}$. This confirms that the analytic reconstruction strategy remains stable with respect to numerical accuracy. Notice noised data points are randomly selected. Other configurations with the same $\rho$ and $n$ may yield even better improvement.
At the same time, higher precision leads to a more sensitive pole detection stage, which typically increases the number of iterations required for convergence. As a result, while the qualitative behaviour is preserved, the computational cost increases with precision. We further extend the analysis in the next section to more realistic scenarios, including datasets affected by both statistical noise and systematic errors, in order to further assess the robustness of the method.

%%%%%%%%%%%%%%%%%%%%%%%%%%%%%%%5
\section{Analysis in a Realistic Scenario: Datasets with Statistic and Systematic Uncertainties}
\label{sec:Analysis in a Realistic Scenario}
In the previous sections, the Padé reconstruction algorithm has been introduced and tested using datasets containing a finite subset of noised data points, showing a good ability to identify and remove them. Real experimental data, however, may be affected by statistical fluctuations and systematic effects with specific analytic structures. To test the performance of the method in a more realistic setting, we consider datasets with statistic and systematic uncertainties, the latest in two representative examples: a Gaussian distortion representing a localized deformation of the observable with no physical analytic structure, and a Breit-Wigner-type contribution, associated with a new resonance in a scattering process and hence singularities in the complex plane. Our goal now is to show explicitly that while the algorithm is able to identify and reduce the systematic uncertainties understood as the aforementioned localized distortions but not the physically meaningful new structures such as a Breit-Wigner distribution, the pure statistical properties of the experimental measurement rest untouched.

For this purpose, in the following two subsections we consider datasets generated from a function defining a physical observable $F(x)$ where another function $G(x)$ has been added. This latter function could be either a meaningless Gaussian distortion in subsection \ref{Gaussian} or a meaningful Breit-Wigner signal in \ref{BW}. In both cases, the starting point is the logarithm function
\begin{equation}
F(x)=\frac{\log(1+x)}{x},
\end{equation}
\noindent
which for example could be interpreted as a scattering cross-section background. 
Data generation is described in Appendix \ref{App:DataGeneration} and emulated the statistical fluctuation of a measurement using a Poisson statistical distribution.

The performance of the analytic reconstruction procedure is also limited by the statistical significance and extent of the distortion. In the present study, the Gaussian distortions that were successfully reconstructed correspond to pulls ranging from approximately $3$ to $30$ and extend over $1$ to $9$ data points. For the Breit--Wigner distribution case, the tested distortions span pulls from $5$ to $50$ and extend over $1$ to $11$ points. These ranges should not be interpreted as absolute limits of the method, but rather as the range explored in the present analysis.

\subsection{Gaussian Distortion}\label{Gaussian}

For this example, to the logarithmic function $F(x)$ an additional contribution $G(x)$ is added as a systematic distortion. It represents a Gaussian term, i.e., a smooth localized deformation which does not introduce additional poles or singularities. In this case, the function $H(x)$ used to generate the pseudo-data is then:

\begin{equation}
H_a(x)=L \left( F(x)+\frac{6}{64}
\exp\left[
-\frac{\left(x-\frac{17}{4}\right)^2}
{5\left(\frac{1}{8}\right)^2}
\right] \right)
\end{equation}

\begin{equation}
H_b(x)=L \left( F(x)+\frac{10}{64}
\exp\left[
-\frac{\left(x-\frac{17}{4}\right)^2}
{5\left(\frac{1}{3}\right)^2}
\right] \right)
\end{equation}

\noindent
The Gaussian contribution is centered at $x=17/4$ and the luminosity factor $L$ is fixed to $L=1000$ in both cases. For the data generation from $H(x)$, the interval $[0,10]$ is divided into $N=30$ bins, resulting in a bin width of $\Delta x_i=1/3$. The observed event counts $N_i$ for each bin centered at $\bar{x}_i$ are generated and subsequently used to reconstruct the corresponding observables $y_i =H(\bar{x}_i)$. An ensemble of 200 pseudo-datasets is then produced by Gaussian sampling at each bin center $\bar{x}_i$, taking $y_i$ as the central value and $\sigma_i$ as the associated standard deviation. Details of data generation and the bootstrapping method followed can be found in Appendix \ref{App:DataGeneration}. 

\begin{figure}[h!]
\centering
\begin{subfigure}[t]{0.48\linewidth}
    \centering
    \includegraphics[width=\linewidth]{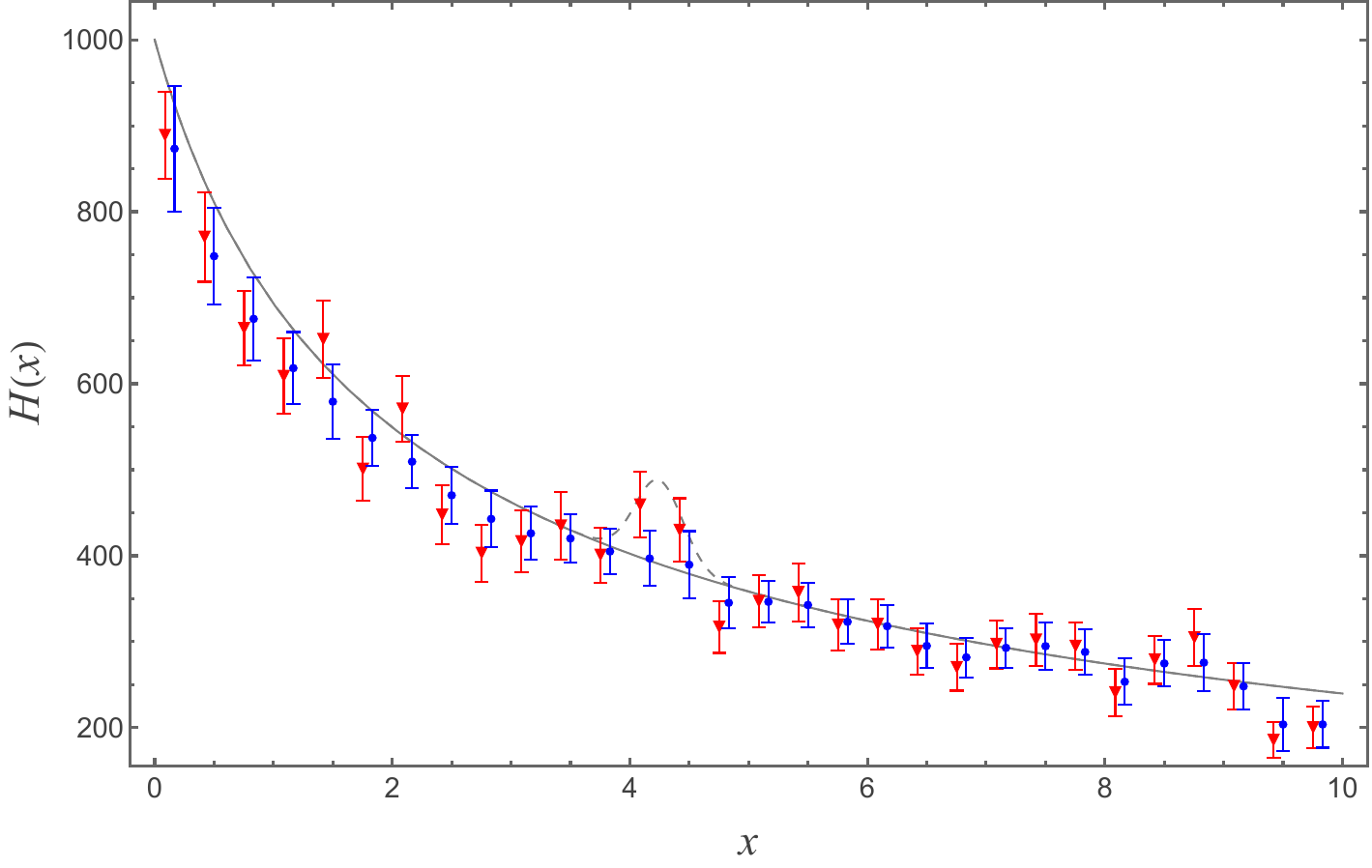}
    \caption{$H_a(x)$ with a small Gaussian distortion of 2 points.}
    \label{Fig:gaussian-small}
\end{subfigure}
\hfill
\begin{subfigure}[t]{0.48\linewidth}
    \centering
    \includegraphics[width=\linewidth]{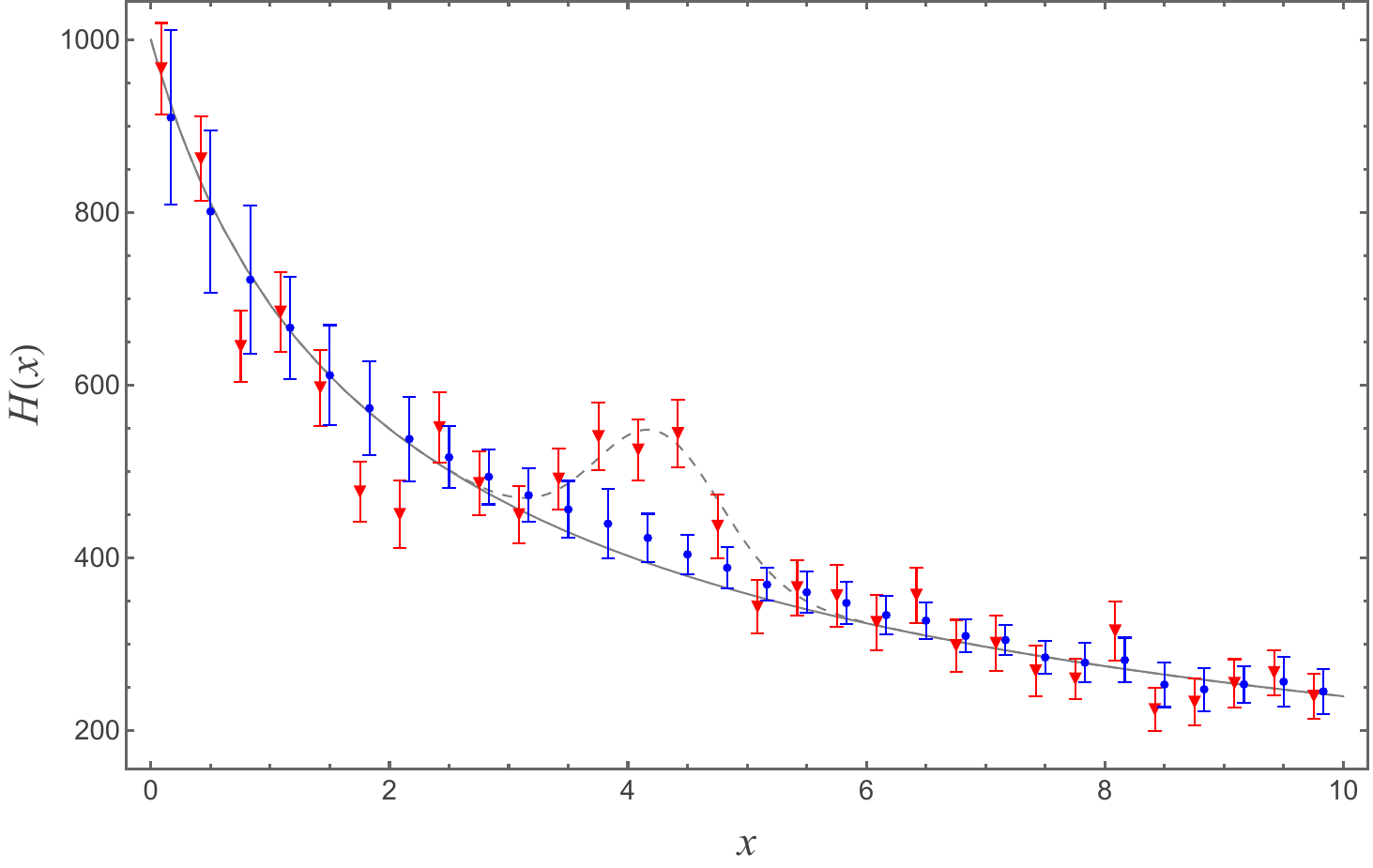}
    \caption{$H_b(x)$ with a large Gaussian distortion of 5 points.}
    \label{Fig:gaussian-big}
\end{subfigure}

\caption{Comparison between the original dataset (red triangles) and the reconstructed dataset (blue dots), including their corresponding uncertainties. For both the small Gaussian distortion $H_a(x)$ and the large Gaussian distortion $H_b(x)$, the analytic procedure successfully reconstructs the distortion, resulting in a closer agreement with  target distribution $F(x)$. For reference, the central value of the initial function $H(x)_{a,b}$ is shown by the gray-dashed curve while the reconstructed one is shown by the solid-gray curve.}
\label{Fig:gaussian}
\end{figure}

The analytic reconstruction procedure is applied to every ensemble, and the resulting distributions are presented in Fig.~\ref{Fig:gaussian} for both $H_a(x)$ and $H_b(x)$ cases. The original 200 peudo-datasets are represented statistically at each $\bar{x}_i$ as triangle points $y_i$ with statistical uncertainty $\sigma_i$. Blue dotes represent the analytically reconstructed data which keep the same statistical uncertainty $\sigma_i$. For easy of comparison, each blue dot is slightly sifted to the right with respect to its red triangle counterpart. Dashed- and solid-gray curves represent the before- and after-reconstruction function $H(x)_{a,b}$. In both cases the gaussian bump at around $x=4$ is clearly seen (dashed representation) and removed (solid representation). Gaussian distortion do not introduce singularities and the algorithm is able, within the statistical uncertainties, to resolve it, distinguishing between $F(x)$ and $H_{a,b}(x)$. We remark, nonetheless, that $H_a(x)$ case is at the edge of resolution of our algorithm as a pull below 3 cannot be distinguished from pure statistical uncertainty. Notice as well how other data points get slightly shifted, specially in the $H_b(x)$, always within the statistically well defined uncertainty.

The refined dataset (in blue) exhibits a clear suppression of the localized Gaussian distortion, resulting in a reconstructed curve that follows the logarithmic function $F(x)$ more closely than the original function $H(x)$. The algorithm is able to identify the Gaussian distortion since it does not introduce a new analytic pattern. As a consequence, the distortion is identified as a spurious contribution and is progressively suppressed during the reconstruction procedure. The resulting improvement indicates that the method is capable of isolating smooth systematic deformations while preserving the analytic structure associated with the physical observable.

To quantify the effect of the analytic reconstruction procedure, we consider the root-mean-square (RMS) uncertainty,

\begin{equation}
\Delta y_{\mathrm{RMS}}
=
\sqrt{
\frac{1}{N}
\sum_{i=1}^{N}
(\Delta y_i)^2
}.
\end{equation}

together with the root-mean-square error (RMSE), 
\begin{equation}
\mathrm{RMSE}
=
\sqrt{
\frac{1}{N}
\sum_{i=1}^{N}
\left(y_i-F(x_i)\right)^2
}
\end{equation}
\noindent

The pull is also considered to quantify the deviation of the reconstructed data from the reference function in units of the corresponding uncertainty,

\begin{equation}
\mathrm{Pull}_i
=
\frac{y_i-F(x_i)}{\Delta y_i}.
\end{equation}

\noindent

After applying the analytic reconstruction procedure, the RMSE with respect to $F(x)$ decreases from $39.64$ to $27.02$ in case (a), corresponding to the small Gaussian distortion, while in case (b), with the larger Gaussian distortion, it decreases from $60.17$ to $15.58$. In both cases, the reduction in RMSE indicates a substantial decrease in the deviation of the reconstructed data from the underlying logarithmic function. The RMS uncertainty, nonetheless, remains essentially unchanged in case (a), changing from $34.57$ to $34.24$, while in case (b) it increases from $35.57$ to $42.57$. The increase observed in case (b), together with the nearly unchanged RMS uncertainty in case (a), indicates that the reduction in RMSE is not simply a consequence of reducing the statistical fluctuations.

The pull also decreases significantly in both cases, from $2.8$ to $0.2$ in case (a) and from $11.2$ to $2.8$ in case (b), indicating that the reconstructed data become considerably more consistent with the underlying function $F(x)$. The combined behavior of these quantities provides evidence that the Gaussian contribution is effectively suppressed by the analytic reconstruction procedure in both cases, while the statistical spread of the reconstructed realizations are not artificially reduced. This result is consistent with the interpretation of the Gaussian contribution as a systematic/noisy deformation of the underlying observable rather than a persistent physical structure.

\subsection{Breit-Wigner Distribution}\label{BW}

For the following case of study, the addition function $H_{a,b}(x)$ is chosen to have a pair of complex-conjugate poles via defining a Breit-Wigner distribution centered at $x=4.25$ and with different "total decay width". Although the resulting functions $H_{a,b}(x)$ are smooth and continuous for all real values of $x$, their analytic structure contains singularities located at the complex variable $z_a = \frac{17}{4} \pm \frac{7}{32}\, i
$ and $z_b = \frac{17}{4} \pm \frac{7}{12}\, i
$ for $H_{a,b}$, respectively. Consequently, such contributions represent a resonance-like structure, a signal with "physical meaning" which should be retained. Since their are well-defined anaylyc structures, we expect the analytic reconstruction algorithm here proposed identify and retain them. The underlying logarithmic function $F(x)$ is the same as in the previous example. The distributions used to generate the pseudo-data are given by
\begin{equation}
H_a(x)=L\cdot \left(\frac{\log(1+x)}{x}
+\frac{0.0011\,x}
{\left(x-\frac{17}{4}\right)^2+\left(\frac{14}{64}\right)^2} \right),
\end{equation}
\begin{equation}
H_b(x)=L\cdot \left(\frac{\log(1+x)}{x}
+\frac{0.012\,x}
{\left(x-\frac{17}{4}\right)^2+\left(\frac{35}{60}\right)^2} \right)
\end{equation}
The statistical setup is identical to that used in the previous example. The luminosity factor is fixed to $L=1000$, and the interval $[0,10]$ is divided into $N=30$ bins of width $\Delta x=1/3$. Following the procedure described above, the expected event counts are computed for each bin and used to generate an ensemble of 200 pseudo-datasets.

\begin{figure}[h!]
\centering

```
\begin{subfigure}[t]{0.48\linewidth}
    \centering
    \includegraphics[width=\linewidth]{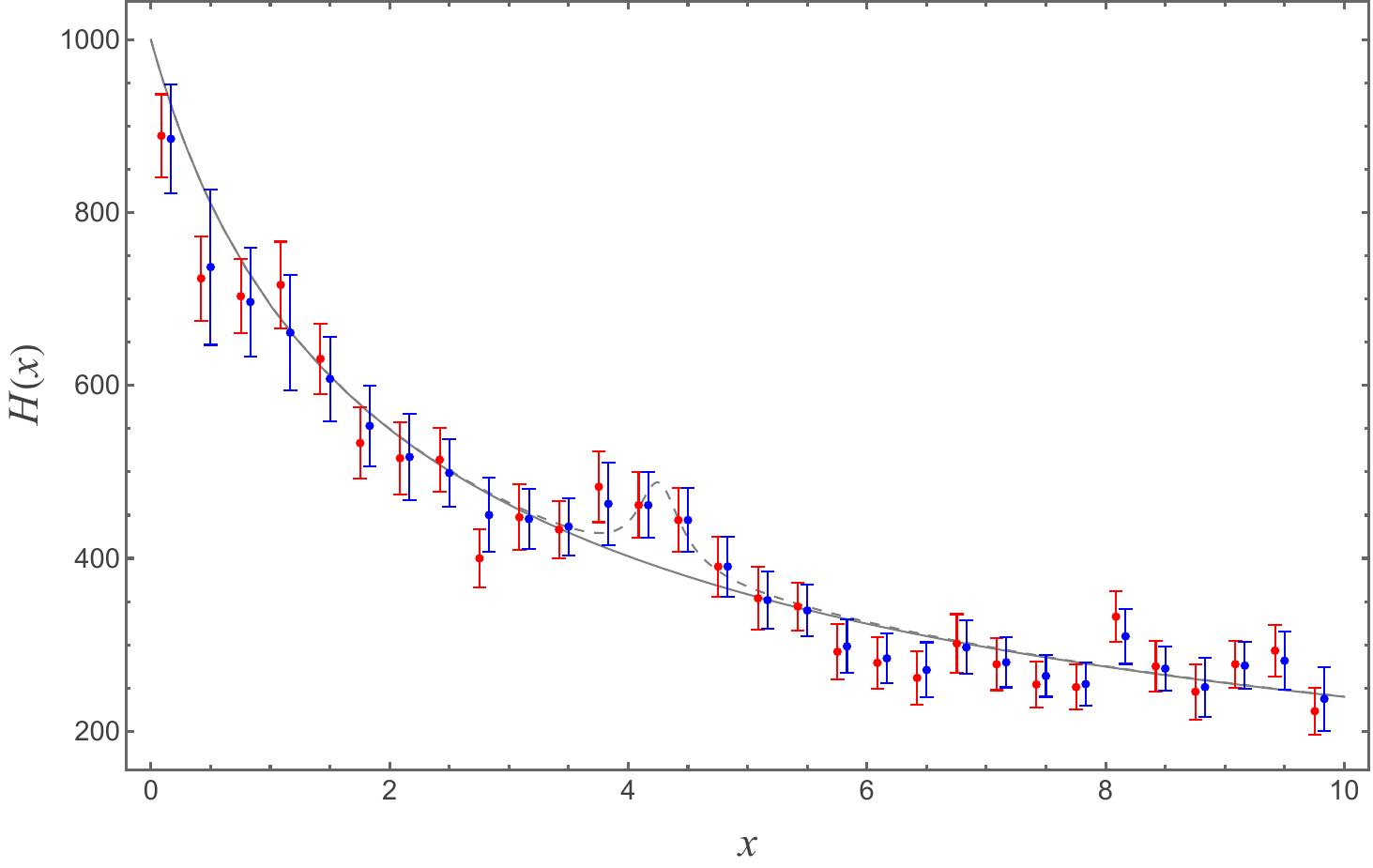}
    \caption{Small Breit-Wigner distortion of 2 points.}
    \label{Fig:BW-small}
\end{subfigure}
\hfill
\begin{subfigure}[t]{0.48\linewidth}
    \centering
    \includegraphics[width=\linewidth]{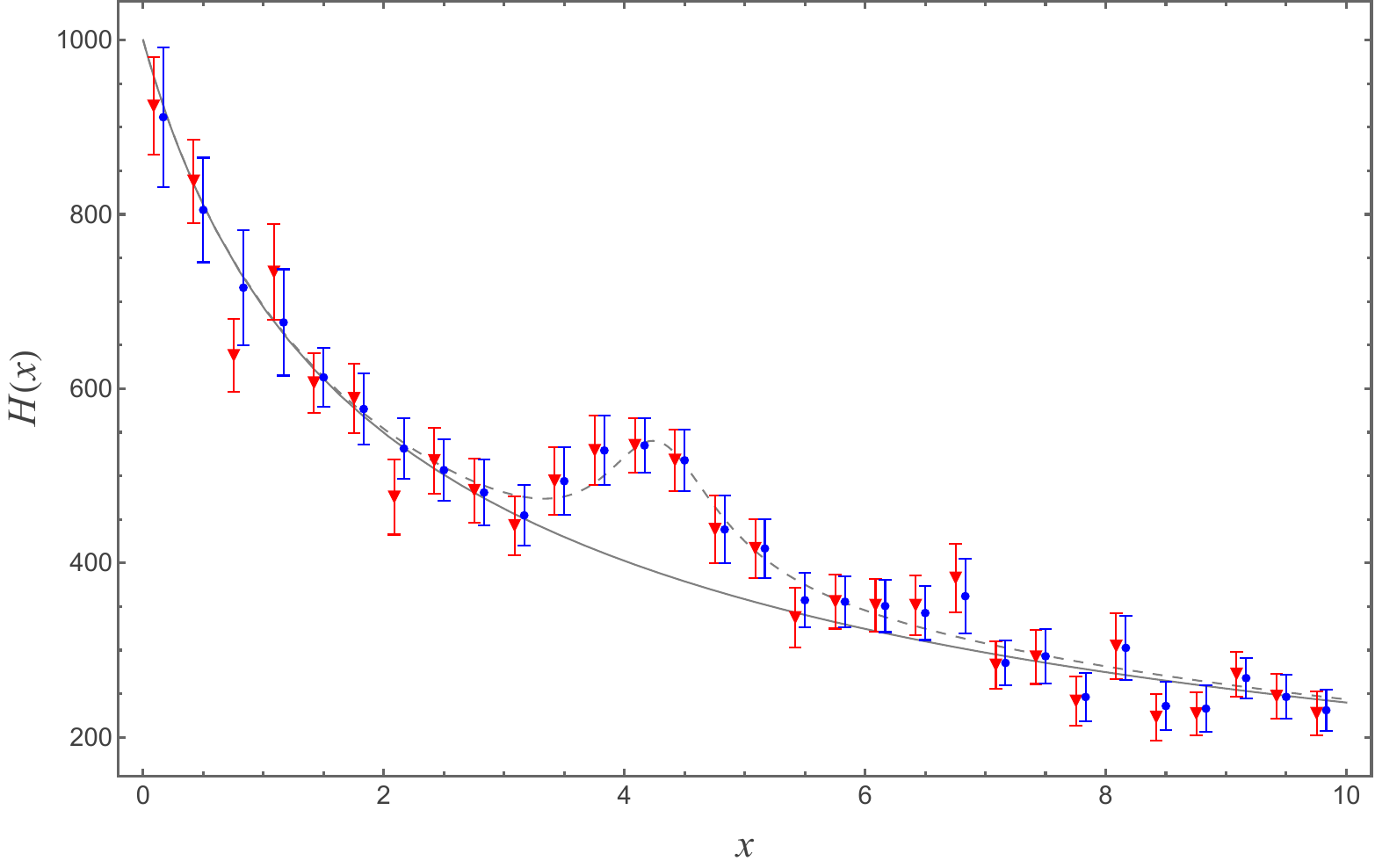}
    \caption{Big Breit-Wigner distortion of 6 points.}
    \label{Fig:BW-big}
\end{subfigure}

\caption{Original data (red triangles) and analytically reconstructed data (blue points) with their corresponding statistical uncertainties. In both cases, the Breit-Wigner resonance-like structure remains visible after the reconstruction procedure is employed. The algorithm also distinguished between Breit-Wigner distribution and logarithmic background as distinguished by the corresponding dashed-gray and solid-gray lines.}
\label{Fig:BW}

\end{figure}

Figure \ref{Fig:BW} summarizes the ability of the algorithm to reconstruct randomly noisy data while keeping all significant physical attributes of the original function which, in the current example, consist of a Breit-Wigner distribution plus a logarithm, or in other words, a resonance-like structure on top of a scattering background. Color criteria is the same as in Figure \ref{Fig:gaussian}: red triangles are generated by the full function $H_{a,b}(x)$ and blue points after the reconstruction algorithm is used.

In case of Fig. \ref{Fig:BW-small}, the RMS statistical uncertainty increases from 35.79 to 42.26, while the RMSE decreases from 40.40 to 31.36, indicating a moderate improvement in the overall reconstruction while preserving the resonant behavior. The pull decreases only slightly from 5.4 to 4.7, indicating that the reconstructed data do not become consistent with the smooth background function within the assigned statistical uncertainties. In case of Fig. \ref{Fig:BW-big}, the RMS statistical uncertainty changes only slightly, from 36.39 to 39.39, while the RMSE decreases from 57.89 to 51.00. The pull remains unchanged at 16.9, providing clear evidence that the reconstructed data are not becoming more consistent with the smooth background function.

The persistence of the resonant contribution, together with the relatively unchanged RMS statistical uncertainty and the large pull values, is consistent with the interpretation of the Breit--Wigner structure as a genuine component of the underlying observable rather than a statistical or systematic distortion. Thus, the analtic reconstruction procedure improves the overall description while preserving the resonance contribution in both cases. In a realistic experimental analysis, such behavior would provide additional confidence that the observed structure is associated with a genuine physical signature rather than with a systematic effect introduced during the measurement process.

\subsubsection{Extraction of the Logarithmic Contribution from the Resonance}

The algorithm here described is so flexible and robust that can be used, alternatively, to separate background from signal of resonance-distribution description. In this subsection, we separate the logarithmic contribution from the resonant structure contained in the dataset. To achieve this, only the part of the data, corresponding to the underlying function $0 < x < 2.5$ and $7.2 < x <10$, is considered. This region is dominated by the background (logarithmic) behavior and is therefore suitable for reconstructing the underlying non-resonant contribution.

\begin{figure}
    \centering
    \includegraphics[width=0.8\linewidth]{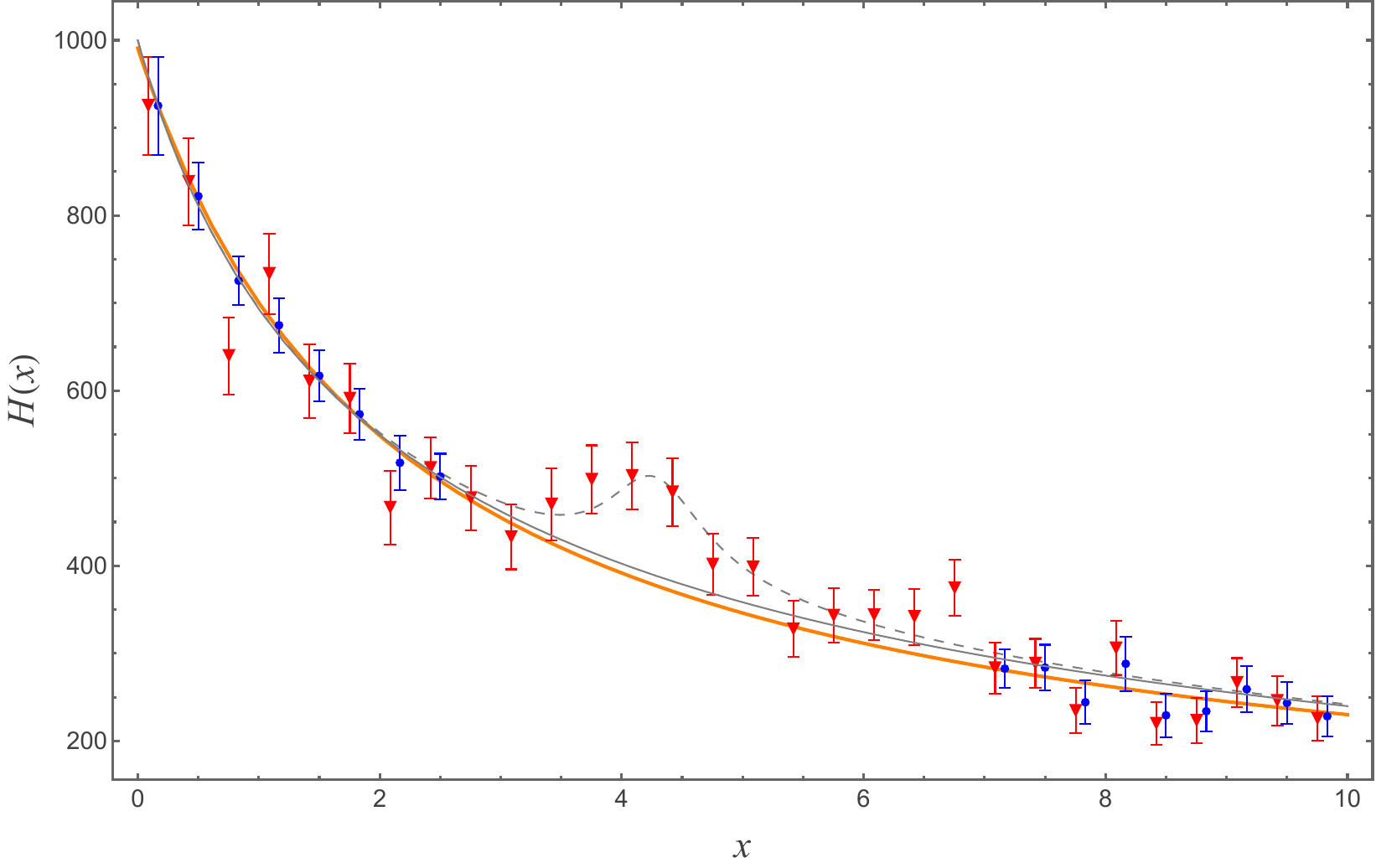}
     \caption{Application of the analytic reconstruction algorithm to subtract background (logarithmic) contribution. Only the data points ($0<x<2.5$) and ($7.2<x<10$) are used to construct a $P_1^1$ (orange curve). Red points represent the original generated pseudo-datasets, while blue points correspond to the reconstructed values. The full distribution $H(x)$ is shown in gray.}\label{Fig:log}
\end{figure}

This selected points are reconstructed using the procedure described in the previous sections. The refined values are then fitted with a $P_1^1$, which is used in turn to reconstruct the logarithmic component of the observable in the full data regime. The resulting approximation is shown in Fig.~\ref{Fig:log} as the orange curve.

\begin{figure}
    \centering
    \includegraphics[width=0.8\linewidth]{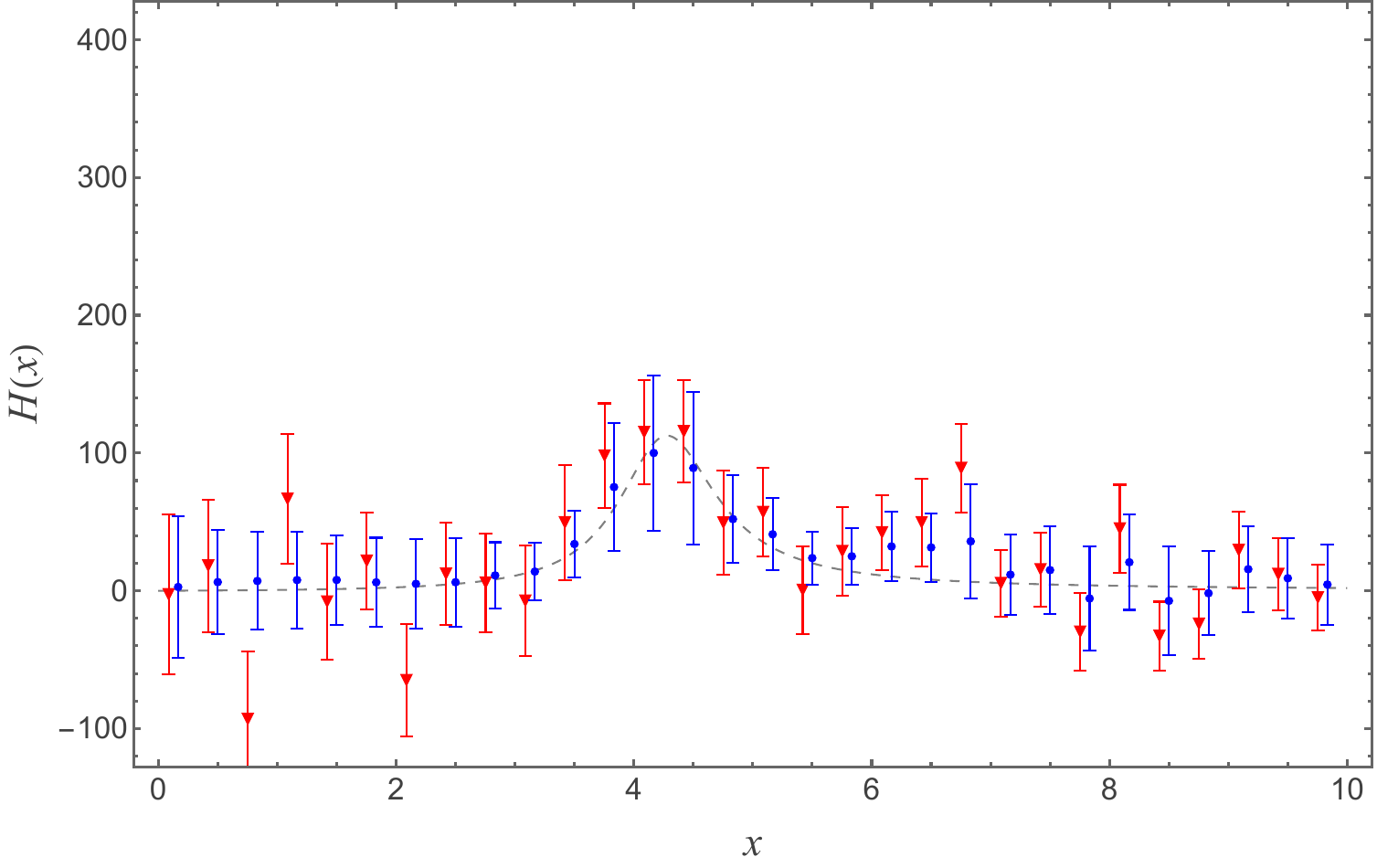}
     \caption{Extraction of the Breit-Wigner resonance after subtraction of the background (logarithmic) contribution. Red triangles denote the original residual data and blue points the reconstructed data. The resonance structure is preserved by the analytic reconstruction procedure, reflecting the stability of its complex-conjugate poles across the Padé sequences. The gray curve corresponds to the original Breit-Wigner contribution.}\label{Fig:filteredBW}
\end{figure}

Once the logarithmic contribution has been reconstructed, its value is removed from each measured point $y_i$. The residual datasets are subsequently processed using the reconstruction algorithm and the resulting distributions reveal the presence of the Breit--Wigner resonance as the dominant remaining structure, as captured in Fig.\ref{Fig:filteredBW}. 

Again the algorithm can isolate the Breit--Wigner distribution thanks to its unique analytic structure which is recurrently reproduced by the PA sequence with its stability indicates not being a random noise. PAs are particularly sensitive to such singularities (complex-conjugated poles, rather than Froissart doublets) and tend to reproduce them consistently across different approximation orders. As a consequence, the resonance is identified as a genuine analytic structure instead of a spurious deformation or a systematic error. The persistence of the Breit--Wigner contribution after the subtraction of the logarithmic background therefore provides additional evidence that the analytic reconstruction procedure is capable of distinguishing resonant structures from non-resonant contributions.

%%%%%%%%%%%

\section{Extension to Holomorphic Functions}
\label{sec:Extension to holomorphic functions}

In this final section, we introduce a straightforward generalization of the proposed method to holomorphic functions. This generalization is particularly useful, as it significantly increases the range of applicability of our analytic reconstruction algorithm, allowing the same noise reconstruction principles to be applied in a broader class of problems beyond the Stieltjes framework.

The modification concerns the definition of the signal-noise separation within a given PA \(P_N^N(x)\) (Eq.~\ref{ec: PAStieltjesNoise}). Since no true singularities are expected within the sampling domain, any real pole appearing in this region is interpreted as a manifestation of noise or data inconsistencies, reflecting a loss of analytic stability in the reconstruction. These contributions are then identified as the noise component \(P_{N-M}^{\,N-M-1}{}_{\mathrm{Noise}}(x)\), while the remaining part of \(P_N^N(x)\) is associated with the holomorphic contribution of the underlying function, the \(P_M^M{}_{\mathrm{Holomorphic}}(x)\).

In addition to this local classification, the decomposition of the PA may also involve global pole structures, such as complex-conjugate pairs. These structures are assigned to the noise or holomorphic components in accordance with their consistency and stability with the global analytic behaviour encoded by the Padé sequence. This point will become clearer in the examples discussed below.

In the following, we illustrate the performance of this generalized procedure through a detailed analysis of noisy discrete data generated from the holomorphic function $(x+1)^{3/2}$. Additional holomorphic examples with different analytic structures are considered in a subsequent subsection, providing further evidence of the robustness of the method beyond the Stieltjes case.

\subsection{Non-Stieltjes Branch-point Example}

To assess the applicability of the proposed reconstruction strategy beyond the Stieltjes class, we consider 
\begin{equation}
g(z) = (z+1)^{3/2},
\label{ec:nonStieltjesbrachpointEx}
\end{equation}
with a branch point at $z=-1$ and a branch cut along the negative real axis. As in the previous sections, noisy discrete datasets are generated from uniform samples of the function on the real axis. In this case, we restrict the analysis to a fixed sampling density $\rho = 2.5$, and consider three noise configurations corresponding to $n=5$, $n=15$, and $n=25$ noisy data points, with relative noise amplitude of up to $20\%$.

%%%%%%%%%%%%%%%%%%%%%%%%%%%%%%%%%%%%%%%%%%%%
\begin{figure}[h!]
\centering

\begin{subfigure}{0.49\textwidth}
    \centering
    \includegraphics[width=\textwidth]{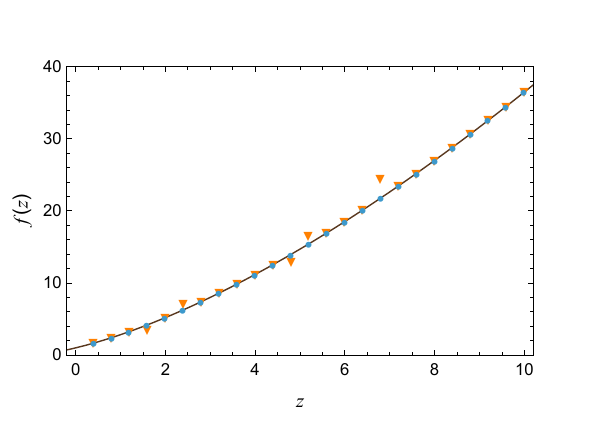}
    \caption{Subset with $\rho = 2.5$ and 5 noisy data points}
\end{subfigure}
\hfill
\begin{subfigure}{0.49\textwidth}
    \centering
    \includegraphics[width=\textwidth]{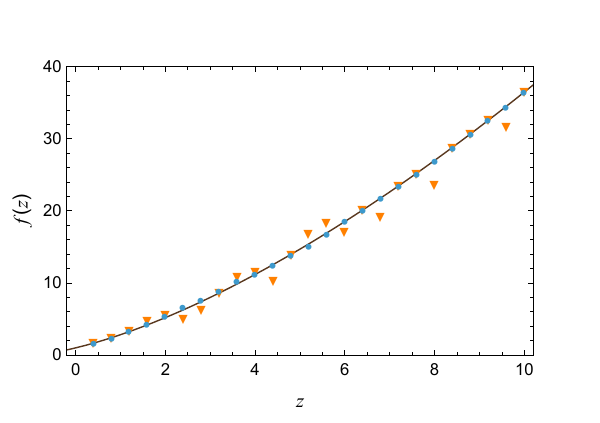}
    \caption{Subset with $\rho = 2.5$ and 15 noisy data points}
\end{subfigure}
\hfill
\begin{subfigure}{0.49\textwidth}
    \centering
    \includegraphics[width=\textwidth]{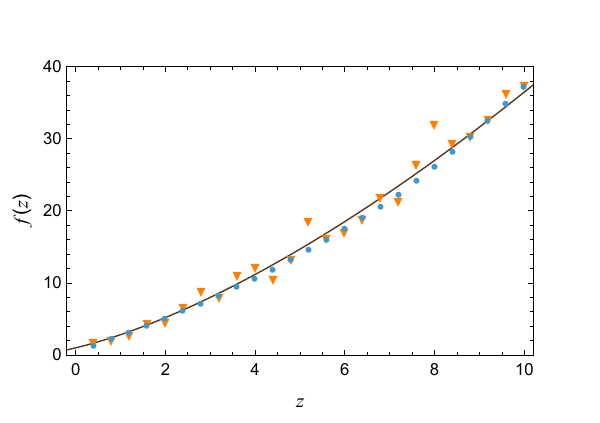}
    \caption{Subset with $\rho = 2.5$ and 25 noisy data points}
\end{subfigure}

\caption{
Representative data subsets of the holomorphic function in Eq.~\ref{ec:nonStieltjesbrachpointEx}, generated under different noise configurations at fixed sampling density $\rho = 2.5$, with numerical precision $10^{-4}$. The black continuous curve corresponds to the exact analytic function. Orange triangles denote the input data after randomly perturbing $n=5,15,25$ sampling points with additive noise of amplitude up to $20\%$. Blue dots represent the reconstructed output produced by the Padé-based algorithm.}
\label{fig:Exampleswithprecision10_4NonStieltjes}
\end{figure}
%%%%%%%%%%%%%%%%%%%%%%%%%%%%%%%%%%%%%%%%%%%%%

In this case, the noise component $P_{N-M}^{\,N-M-1}{}_{\mathrm{Noise}}(x)$ is defined as follows: real poles within the sampling region are attributed to noise, while complex-conjugate structures are also included in the noise contribution as they are not compatible with the analytic behavior of $g(z)$.

 The results are shown in Fig.~\ref{fig:Exampleswithprecision10_4NonStieltjes} and summarized in Table~\ref{tab:MAE_filtering_Ex2}. In all cases, the analytic reconstruction procedure leads to a consistent reduction of the reconstruction error, yet slightly smaller than in the Stieltjes case. The improvement is most pronounced in the low-noise regime, but remains clearly visible even in the most strongly perturbed cases.

%%%%%%%%%%%%%%%%%%%%%%%%%%%%%%%%%%%%%%%%%%%%5
\begin{table}[]
\caption{MAE before and after applying the reconstruction algorithm, together with the corresponding relative improvement for different noise configurations at a fixed sampling density $\rho = 2.5$. The results correspond to the data subsets shown in Fig.\ref{fig:Exampleswithprecision10_4NonStieltjes}.}
\label{tab:MAE_filtering_Ex2}
\begin{tabular}{ccccc}
\hline
                            & Total Number of   & MAE              & MAE                 & Relative           \\
                            & Noisy Data Points & Before Reconstruction & After Reconstruction     & Improvement $(\%)$ \\ \hline
\multirow{3}{*}{$\rho=2.5$} & 5                 & $0.26$           & $8.0 \cdot 10^{-6}$ & $100$              \\
                            & 15                & $0.82$           & 0.11              & $86.84$            \\
                            & 25                & $1.1$            & $0.48$              & $56.94$            \\ \hline
\end{tabular}
\end{table}
%%%%%%%%%%%%%%%%%%%%%%%%%%%%%%%%%%%%%%%%

\subsection{Additional Holomorphic Test Functions}

Beyond the branch-point example discussed above, we have applied the reconstruction procedure to other holomorphic functions, including $e^x$, $\tan^2(x)$, and $\log^2(1+x)/x^2$. In all cases, the method provides a consistent identification of noise contributions from the underlying analytic structure, and reduces the systematic uncertainties.

A particularly instructive case is provided by $e^x$. In this setting, the PAs constructed from its Taylor expansion generate sequences of poles that, as the approximation order increases, accumulate along a curve in the complex plane, commonly referred to as a Szegő-type curve \cite{baker1996pade}. In this way, complex structures associated with the Szegő curve are naturally incorporated into \(P_M^M{}_{\mathrm{Holomorphic}}(x)\), while \(P_{N-M}^{\,N-M-1}{}_{\mathrm{Noise}}(x)\) is restricted to real poles within the sampling domain. 

%%%%%%%%%%%%%%%%%%%%%%%%%%%%%%%%%%%%%%%%%%%%%%%%%%%%%
%%%%%%%%%%%%%%%%%%%
\section{Conclusions}
\label{sec:conclusion}

We have introduced a new perspective on the role of PAs in the analysis of finite datasets. Rather than viewing Froissart doublets solely as undesirable numerical artifacts, we have shown that they constitute \textit{diagnostic objects} whose dynamics provide direct information about the analytic consistency of the underlying data. This reinterpretation forms the basis of a general reconstruction algorithm that exploits the evolution of Padé sequences to separate genuine analytic information from inconsistencies introduced by finite-information effects.

The proposed method differs fundamentally from conventional filtering or smoothing techniques. Its objective is not to suppress fluctuations according to statistical criteria, but to identify those components of the dataset that are incompatible with a common analytic structure. Consequently, the algorithm naturally preserves statistical fluctuations that remain consistent with the expected analytic behavior, while selectively identifying and correcting localized systematic distortions that manifest themselves as analytic inconsistencies. By relying exclusively on analytic properties of PAs, the algorithm remains largely independent of the physical origin of the inconsistencies, making it applicable to statistical fluctuations, localized systematic distortions, finite numerical precision, or other sources of imperfect information.

The numerical examples demonstrate that the method successfully reconstructs the underlying analytic structure while preserving genuine physical singularities and the statistical fluctuations naturally associated with finite datasets. At the same time, it selectively suppresses localized systematic distortions that are incompatible with the analytic structure of the underlying function. This ability to discriminate between statistical variability and systematic analytic inconsistencies distinguishes the proposed approach from conventional denoising strategies. 

Although our examples focus on Stieltjes and holomorphic functions, the underlying philosophy is considerably broader. We expect that exploiting the analytic consistency encoded in Padé sequences may provide a useful framework whenever one seeks to infer reliable analytic information from finite datasets.

Finally, we have emphasized the practical character of the method. To encourage its use in other applications, the complete implementation is distributed as as a supplementary
Mathematica notebook in an open \textcolor{red}{\href{https://gitlab.pic.es/bduch/reconstruction-finite-data-pade-sequences}{GitLab}} repository. We hope that this work will stimulate further developments in which Padé approximants are employed not only as tools for approximation or analytic continuation, but also as quantitative probes of the analytic consistency of experimental and numerical data.

%%%%%%%%%%%%%%%%%%%%%%%%%%%%%%%%%%%%%%%%%%%%%%%%%%%%%
\begin{appendix}
\label{app}
\section{Supplemental Material for the Analytically Consistent Reconstruction Algorithm: Conventions}
\label{app:Conventions}

The numerical analysis has been implemented in \texttt{Mathematica}. The corresponding code is provided as supplementary material. A reference implementation of the algorithm, together with illustrative examples, is also available in the associated \textcolor{red}{\href{https://gitlab.pic.es/bduch/reconstruction-finite-data-pade-sequences}{GitLab}} repository. In this Appendix, we summarize the conventions, main functions, and tunable parameters used in the implementation

\subsection{Main functions of the Analytically Consistent Reconstruction Algorithm}

The  algorithm is structured around three interdependent modules that define the Padé-based reconstruction, the stochastic correction step, and the iterative refinement procedure.
\begin{enumerate}
    \item \begin{verbatim}
PAOriginalPAStieltjesPANoiseSplit[n, m, data]
\end{verbatim} This function performs a PA of order \((M,N)\) on the input dataset and returns the reconstructed function together with its decomposition into Stieltjes (holomorphic in the generic case) \(P_M^M{}_{\mathrm{Stieltjes}}(x)\) and noise components \(P_{N-M}^{\,N-M}{}_{\mathrm{Noise}}(x)\). It also provides diagnostic information on the effective Padé orders of both components.

    \item \begin{verbatim}
UniversalNonEquidistantCorrectionFunctions[data, targetX, StieltjesFunction]
\end{verbatim}
This routine performs local stochastic corrections of selected data points guided by the analytic component of the Padé reconstruction. For each target position $x_i$, the reference value is given by the corresponding evaluation of the Stieltjes component, which defines the local analytic constraint for the update. Each affected point is updated by sampling from a Gaussian distribution centered at the analytic prediction, with a truncation scheme that depends on the relative deviation of the original value. The procedure is iterated until a local smoothness criterion with respect to nearest neighbors is satisfied. This incorporates a discrete convexity requirement, which reduces in the Stieltjes case to a non-negative second derivative in the continuum limit. In the more general holomorphic case, this is replaced by the corresponding analytic consistency condition. When strict enforcement is obstructed by local noise, the constraints are relaxed to bounded deviations, allowing controlled departures within the intrinsic numerical precision of the dataset.

\item \begin{verbatim}
RefineDataByPadeAnalysisStieltjesNonEquid[inputData]
\end{verbatim} 
This routine implements a full iterative refinement algorithm for non-equidistant datasets based on Padé analysis. At each iteration, a diagonal PA of order $(N,N)$ are constructed using \texttt{PAOriginalPAStieltjesPANoiseSplit} over a range of truncation orders, with the request to probe the stability of the reconstruction. The pole structure of the noise component is then analyzed to identify spurious singularities, which are interpreted as indicators of locally inconsistent data points. A statistical voting scheme is used to cluster poles and associate them with  anomalous points candidate in the dataset. In parallel, the Stieltjes component (holomorphic component in the generic case) is selected by minimizing the reconstruction error across different Padé orders, providing a stable analytic reference function. Once the reference Stieltjes function is determined, data points identified as anomalous are updated using the local stochastic correction procedure implemented in the previous routine \texttt{UniversalNonEquidistantCorrectionFunctions}, which enforces consistency with the reconstructed analytic structure. The procedure is iterated until convergence, defined by the absence of persistent poles, negligible successive corrections within numerical precision, or the suppression of leading vote falls below a prescribed threshold (including the limiting case of zero threshold).
\end{enumerate}

\subsection{Model specification and parameter structure of the algorithm}

Given a particular analysis, the algorithm contains certain hardwired parameters that must be adapted to the specific dataset under study. The numerical implementation depends on a set of adjustable parameters which control the stability, resolution, and convergence properties of each stage of the algorithm. The variation of these parameters do not affect the qualitative structure of the method. In the following three Tables~\ref{tab:pade_params}-\ref{tab:refinement_full_params}, we summarize the relevant parameter sets, each table associated with one of the main computational components of the algorithm.
%%%%%%%%%%%%%%%%%%%%%%%%%%%%%%%%%%%%%%%%%%
\begin{table}[!h]
\centering
\caption{Parameters controlling \texttt{PAOriginalPAStieltjesPANoiseSplit[n, m, data]}.The symbol $\textsuperscript{*}$ denotes parameters that must be adapted to the specific dataset under study.}
\label{tab:pade_params}
\begin{tabular}{l l p{5.2cm} l}
\hline
Parameter & Description & Role & Default \\
\hline
$n, m$ & Padé orders & Controls approximation complexity & data-dependent \\
WorkingPrecision & Precision & Stability of nonlinear fit & 40$\textsuperscript{*}$ \\
PrecisionGoal & Precision goal & Convergence criterion & 6$\textsuperscript{*}$ \\
AccuracyGoal & Accuracy goal & Fit reliability control & 6$\textsuperscript{*}$ \\
MaxIterations & Iteration limit & Prevents non-convergence & 500$\textsuperscript{*}$ \\
StieltjesCriterion & Pole condition & Defines Stieltjes sector via Padé poles in admissible region & model-dependent \\
\hline
\end{tabular}
\end{table}
%%%%%%%%%%%%%%%%%%%%%%%%%%%%%%%%%%%%%%%%%%%%%%%%%%%%%%%%%%%%%%%%%%%%%%%%%%%%%%%%%%%%
\begin{table}[!h]
\centering
\caption{Parameters controlling \texttt{UniversalNonEquidistantCorrectionFunctions[data, targetX, StieltjesFunction]}. The symbol $\textsuperscript{*}$ denotes parameters that must be adapted to the specific dataset under study.}
\label{tab:correction_params}
\begin{tabular}{l l p{4.2cm} l}
\hline
Parameter & Description & Role & Default \\
\hline
sigma & Gaussian width & Stochastic fluctuation scale & $10^{-5}\textsuperscript{*}$ \\
curvatureLimit & Reference scale & Analytic constraint strength & input function \\
tolerance & Smoothness factor & Update acceptance criterion & $1.0\textsuperscript{*}$ \\
prec & Precision threshold & Final rounding accuracy & $10^{-4}\textsuperscript{*}$ \\
\hline
\end{tabular}
\end{table}
%%%%%%%%%%%%%%%%%%%%%%%%%%%%%%%%%%%%%%%%%%%%%%%%%%%%%%%%%%%%%%%%%%%%%%%%%%%%%%%%%%%%
\begin{table}[!h]
\centering
\small
\caption{Parameters and internal variables controlling \texttt{RefineDataByPadeAnalysisStieltjesNonEquidProof}. The symbol $\textsuperscript{*}$ denotes dataset-dependent or adaptive quantities.}
\label{tab:refinement_full_params}

\begin{tabular}{l l p{3.8cm} l}
\hline
Parameter & Description & Role & Default \\
\hline

diagOrders & Padé order range & Stability scan of reconstruction & $5$--$12\textsuperscript{*}$ \\
prec & Numerical cutoff & Acceptance threshold & $10^{-4}\textsuperscript{*}$ \\
maxConsecutive & Stall detection window & Prevents cycling in iterations & $6\textsuperscript{*}$ \\
vote threshold & Pole multiplicity cutoff & Detection of anomalous points & $>1\textsuperscript{*}$ \\
maxOrd & Model selection cutoff & Optimal Padé/Stieltjes fit selection & adaptive \\

uniqueX & Grid points & Non-equidistant sampling structure & input data \\
nf & Nearest-neighbour map & Local indexing / interpolation tool & constructed \\
localDiffs & Grid spacing & Local adaptive scale definition & computed \\

currentTolerance & Pole acceptance threshold & Adaptive filtering of pole candidates &$ 0.45\,\Delta x\textsuperscript{*}$ \\
voteTally & Pole clustering histogram & Statistical identification of singularities & computed \\
voteLevels & Vote hierarchy structure & Multilevel candidate filtering & derived \\

scores & Mean absolute reconstruction error & Ranking of PAs & computed \\
StieltjesFunction & Selected analytic reconstruction & Reference function for correction step & selected per iteration \\

historyWinners & Convergence memory buffer & Prevents cycling / detects stalls & empty list \\
\hline

\end{tabular}
\end{table}

\section{Data Generation for the Realistic Scenario of Section \ref{sec:Analysis in a Realistic Scenario}}\label{App:DataGeneration}

In this appendix we describe the procedure used to generate the statistical component of the synthetic datasets exploited in Sec.\ref{sec:Analysis in a Realistic Scenario}. The starting point is a function $F(x)$ which represents the underlying physical observable. A second contribution, $G(x)$, is then introduced to model a localized distortion. The combination of both contributions defines the physical observable to be measured.
\begin{equation}
H(x)=F(x)+G(x)
\tag{5}
\end{equation}
The interval $[x_{\min},x_{\max}]$ is then divided into $N_{\mathrm{bins}}$ bins with boundaries $x_i$. Assuming an integrated luminosity $L$, the expected number of events in the $i$-th bin is defined as:
\begin{equation}
\lambda_i = L \int_{x_i}^{x_{i+1}} H(z)\,dz
\end{equation}
This quantity represents the average number of events predicted by the physical model within that bin. Since an experiment does not observe exactly $\lambda_i$, the observed number of events is generated by

\begin{equation}
N_i \sim \mathrm{Poisson}(\lambda_i).
\end{equation}

The measured observable is reconstructed according to

\begin{equation}
y_i = \frac{N_i}{L\,\Delta x_i}.
\end{equation}

The associated uncertainty is defined by Poisson statistics,

\begin{equation}
\sigma_i = \frac{\sqrt{N_i}}{L\Delta x_i}.
\end{equation}

A bin center is assigned to each interval along the $x$-axis,

\begin{equation}
\overline{x}_i = \frac{x_i + x_{i+1}}{2}
\end{equation}

The resulting dataset is therefore represented by the triplets
$(\overline{x}_i,y_i,\sigma_i)$, which contain the bin position, the reconstructed observable, and its statistical uncertainty, respectively. To generate the ensembles used in the reconstruction procedure, $N_{\mathrm{pseudo}}$ pseudo-datasets are subsequently constructed by sampling around each $y_i$ according to its corresponding statistical uncertainty $\sigma_i$,

\begin{equation}
y_{i,k}=y_i+\epsilon_{i,k},
\qquad
\epsilon_{i,k}\sim\mathcal{N}(0,\sigma_i),
\end{equation}

\noindent
where $k=1,\ldots,N_{\mathrm{pseudo}}$. In the present analysis, $N_{\mathrm{pseudo}}=200$. Each pseudo-dataset contains the same bin positions $\overline{x}_i$, while the values $y_{i,k}$ fluctuate according to the statistical uncertainty obtained from the Poisson event counts. No detector smearing or detection-efficiency effects are included in this construction, so that the generated datasets isolate the statistical fluctuations associated with event counting.

\end{appendix}
%%%%%%%%%%%%%%%%%%%%%%%%%%%%%%%%%%%%%%%%%%%%%%%%%%%%%
\section*{Acknowledgment} 
This work has been supported by the Ministerio de Ciencia e Innovaci\'on under grant PID2023-146142NB-I00 and through the State Research Agency under the Severo Ochoa Centres of Excellence Programme 2025-2029 (CEX2024-001442-S), by the Secretaria d'Universitats i Recerca del Departament d'Empresa i Coneixement de la Generalitat de Catalunya under grant 2021 SGR 00649. IFAE is partially funded by the CERCA program of the Generalitat de Catalunya.
B.~D.\ also acknowledges support from the predoctoral program AGAUR-FI (2025 FI-1 00461) Joan Or\'o of the Department of Research and Universities of the Generalitat de Catalunya, co-financed by the European Social Fund Plus. 
E.~D.\ acknowledges support from the Secretaría de Ciencia, Humanidades, Tecnología e Innovación (SECIHTI).

%%%%%%%%%%%%%%%%%%%%%%%%%%%%%%%%%%%%%%%%%%%%%%%%%%%%%
%%%%%%%%%%%%%%%%%%%%%%%%%%%%%%%%%%%%%%%%%%%%%%%%%%%%%

\appendix
%%%%%%%%%%%%%%%%%%%%%%%%%%%%%%%%%%%%%%%%%%%%%%%%%%%%%
%%%%%%%%%%%%%%%%%%%%%%%%%%%%%%%%%%%%%%%%%%%%%%%%%%%%%

%%%%%%%%%%%%%%%%%%%%%%%%%%%%%%%%%%%%%%%%%%%%%%%%%%%%%
%%%%%%%%%%%%%%%%%%%%%%%%%%%%%%%%%%%%%%%%%%%%%%%%%%%%%

\end{document}